\documentstyle[aps,prl,preprint]{revtex} 
 
\draft 
\tighten

\begin{document} 
 
\title{Comment on "First Observation for a Cuprate 
Superconductor of Fluctuation-Induced Diamagnetism Well Inside the 
Finite-Magnetic-Field Regime"}

\author{ 
T. Mishonov\thanks{tel.:(++32)-16-327193, 
                   fax.:(++32)-16-327983, 
                   e-mail:~todor.mishonov@fys.kuleuven.ac.be}%
           \thanks{On leave from: Department of Theoretical Physics, 
                   Faculty of Physics, University of Sofia, 
                   5 J. Bourchier Blvd., Bg-1164 Sofia, Bulgaria, 
                   e-mail:~mishonov@rose.phys.uni-sofia.bg} 
} 
\address{%
 Laboratorium voor Vaste-Stoffysica en Magnetisme, 
 Katholieke Universiteit Leuven,\\ 
 Celestijnenlaan 200 D, B-3001 Leuven, Belgium\\ 
} 
\preprint{Submitted to Phys. Rev. Lett.} 
 
\maketitle 
\ \\ 

The recent Letter by Carballeira {\it et al.}~\cite{Carballeira} is 
an important hint that fluctuation-induced diamagnetism in 
La$_{1.9}$Sr$_{0.1}$CuO$_4$ is due to 
Gaussian fluctuations of the Ginzburg-Landau (GL) order 
parameter. These results show that the accuracy of the GL theory with cutoff 
in the in-plane energy will probably 
be enough for an adequate description of the 
fluctuation phenomena in high-$T_c$ superconductors. 
Surprisingly, I found, however, 
that the authors of Ref.~\cite{Carballeira} 
have used expressions to fit their experimental data identical to the 
formulae of the review Ref.~\cite{Mishonov00} 
where a detailed derivation is presented. 
Indeed, the local two-dimensional (2D) limit ($r\ll \epsilon + h\ll c$) for 
the fluctuation magnetisation Eq.~(3) of Ref.~\cite{Carballeira} 
is identical to Eq.~(145) of Ref.~\cite{Mishonov00}. 
Likewise, the formula for the 2D magnetisation Eq.~(2) of 
Ref.~\cite{Carballeira} is nothing but a special case of the general 
expression Eq.~(135) of Ref.~\cite{Mishonov00} for the fluctuation 
magnetisation of a ``single layered superconductor'', 
\begin{eqnarray} 
  \Delta M(\epsilon,h;r,c)   & = &   -\frac{k_B T}{\Phi_0 S} 
    \left( 
        \frac{c}{2h} + {2\over\pi} \int_{0}^{\pi/2}{\rm d}\varphi 
      \left\{ 
        -\ln\Gamma\left( 
                    \frac{\epsilon+h+r\sin^2\varphi}{2h} 
                  \right) 
      \right. 
    \right.               \nonumber \\ 
    & + & \ln\Gamma\left( 
                    \frac{c + \epsilon + h + r \sin^2\varphi}{2h} 
                  \right) 
    + \left[ 
        \frac{\epsilon + r\sin^2\varphi}{2h} 
        \psi\left( 
            \frac{\epsilon+h+r\sin^2\varphi}{2h} 
            \right) 
      \right. \label{magnetization} \\ 
  & - & 
  \left. 
    \left. 
      \left. 
        \frac{c + \epsilon + r \sin^2\varphi}{2h} 
          \psi\left( 
               \frac{c + \epsilon +h + r \sin^2\varphi}{2h} 
              \right) 
      \right] 
    \right\} 
  \right), \nonumber 
\end{eqnarray} 
where for small values of the Lawrence-Doniach (LD) dimensional crossover 
parameter $r=(2\xi_c(0)/s)^2 \ll\epsilon + h$ the averaging with respect 
to the Josephson phase $2\varphi$ can be omitted. 
At the critical temperature,  $\epsilon=0,$ and high magnetic fields, 
$h\gg r$, the 2D approximation, 
Eq.~(137) of Ref.~\cite{Mishonov00}, predicts an universal scaling 
low given in Fig.~1 of Ref.~\cite{Mishonov00}, 
\begin{equation} 
-\Delta M(\epsilon=0,h\gg r;c)  =  \frac{k_B T}{\Phi_0}\left\{ 
    \ln\Gamma\left({1\over y}+{1\over 2}\right) 
  - {1 \over 2} \ln\pi 
  + \left[ 1 - \psi\left({1\over y}+{1\over 2}\right)\right]\right\}, 
\label{2D} 
\end{equation} 
where $y=2h/c$. In the opposite case of local 3D behavior 
near the phase curve $H_{c2}(T)$ ($\epsilon+h\ll r,c$) 
the fluctuation magnetic moment is expressed via the Hurwitz 
$\zeta$-functions, Eq.~(183) of Ref.~\cite{Mishonov00}, 
\begin{equation} 
-\Delta M(\epsilon,h\ll r,c)  =  \frac{k_B T}{\Phi_0} 
 3\left( {2 \over r} \right)^{1/2} 
 \sqrt{h} \left[ \zeta\left(-{1\over 2},{1\over 2}+{\epsilon\over 2h}\right) 
  - {1\over 3} \zeta\left({1\over 2},{1\over 2} 
  + {\epsilon\over 2h}\right){\epsilon\over 2h} \right]; 
\label{zetaM} 
\end{equation} 
cf. also Ref.~\cite{Mishonov89}, where the $\zeta$-function method for 
ultraviolet regularization has been already applied to the GL theory of 
Gaussian fluctuations for an uniaxial superconductor. 
At the critical temperature, $T=T_c,$ this local formula, 
$h\ll r,c$ gives the well-known 
Prange result with a correction for the anizotropy, 
Eq.~(184) of Ref.~\cite{Mishonov00}, 
\begin{equation} 
-\Delta M(h)  =  \left(\frac{k_B T}{\Phi_0}\right) 3 
      \sqrt{2} \zeta\left(-{1\over 2},{1\over 2}\right) \sqrt{h\over r} 
              =  3 \pi^{1/2} \zeta\left(-{1\over 2},{1\over 2}\right) 
                   {k_B T\over \Phi_0^{3/2}} \frac{\xi_{ab}(0)}{\xi_{c}(0)} 
                   \sqrt{\mu_0 H}. 
\label{Prange} 
\end{equation} 
 
The above findings, thus, strongly point to the ultimate importance of the 
magnetisation curves at $T_c$ as it is well described in the textbook 
by Tinkham\cite{Tinkham}. 
Therefore, instead of only an order evaluation like $c\simeq {1\over2},$ 
a reliable determination of the energy cutoff parameter for CuO$_2$ plane $E_{\rm max}=c 
\hbar^2/2m^*_{ab}\xi_{ab}^2(0)$ and LD parameter $r$ can be achieved 
by a two parameter fit of the magnetisation curve $-\Delta M(T_c, H; r,c)$ 
employing the complete analytical result 
for a layered superconductor Eq.~(\ref{magnetization}). 
The weak magnetic field region (behind the critical region near $h=0$ 
which could be influenced also by the disorder), cf. Eq.~(\ref{Prange}), 
determines the anisotropy parameter 
$\xi_{ab}(0)/\xi_{c}(0)=\left(m^*_{c}/m^*_{ab}\right)^{1/2}$ 
and LD parameter $r,$ 
while the strong magnetic field asymptotics fixes the scale of the $y$ 
variable from Eq.~(\ref{2D}) and the cutoff 
parameter $c$. If those values come to be in agreement with an analogous fit to 
the fluctuation conductivity based upon Eq.~(198) of Ref.~\cite{Mishonov00} 
we could consider the \textit{Gaussian spectroscopy} of high-$T_c$ materials 
for having been already established. The aim of the present 
comment is to provoke the performance of simple additional 
measurements at $T_c$ which, after a professional experimental data 
processing, will reveal an important quantitative information 
concerning the GL parameters and applicability of the GL theory 
for high-$T_c$ superconductors in general. 
Confirmation of this would be crucial for the further 
studies of the fluctuation phenomena in superconductors. 

\begin{references} 
\bibitem{Carballeira} C.~Carballeira, J.~Mosqueira, A.~Revcolevschi, 
                  and F.~Vidal, Phys. Rev. Lett. {\bf 84}, 3157 (2000). 
%
\bibitem{Mishonov00}T.~Mishonov and E.~Penev, 
          \textit{Thermodynamics of Gaussian fluctuations and 
                  paraconductivity in layered superconductors}, 
                  cond-mat/0004023 (To be published as review 
                  article in International Journal of Modern 
                  Physics B). 
%
\bibitem{Mishonov89}T. Mishonov, "Fluctuation Torque for 
                  High Temperature Superconductors in High Magnetic Fields", 
                  in {\it Physics and Materials Science of High Temperature 
                  Superconductors}, 
                  Edited by R. Kossowsky {\it et al.} 
                  (Kluwer Academic Publishers, Dordrecht, 1990), 
                  NATO ASI E \textbf{181}. 
%
\bibitem{Tinkham} M.~Tinkham, {\it Introduction to Superconductivity} 
                 (McGraw-Hill, New York, 1996), Chap. 8, Fig.~8.5, p.~306. 
\end{references}
\end{document}